\documentclass[twocolumn,pre]{revtex4}
\usepackage{graphicx}
\begin{document}

\title
{DNA melting: intra base-pair dynamics and a 
vector
 generalization of the Peyrard-Bishop-Dauxois model}
\author{Nikos Theodorakopoulos$^{1,2}$ 
} 
\affiliation{
$^{1}$Theoretical and Physical Chemistry Institute, National Hellenic Research
Foundation,\\ 
Vasileos Constantinou 48, 116 35 Athens, Greece \\
$^{2}$Fachbereich Physik, Universit\"at Konstanz, 78457 Konstanz, Germany\\
}
\date{\today}
\begin{abstract}
The Peyrard-Bishop-Dauxois (PBD) model of DNA denaturation, although successful in the description of melting profiles, fails to predict melting entropies, unzipping forces and dynamical properties, e.g. hairpin dynamics. The paper presents an atomistic \lq\lq toy model\rq\rq\- of the intra base-pair motion which suggests that the thermodynamics may be better described by a planar vector - rather than a scalar - order parameter.
This leads to correct estimates of melting entropy, unzipping force, hairpin opening rates, and the equilibrium constant of open/closed base pair states during imino proton exchange.
\end{abstract}
\maketitle
\nopagebreak
\section{Introduction}
Mesoscopic modeling is a key tool in the exploration of important biophysical
processes which occur at or above the nanoscale. 
The
Peyrard-Bishop-Dauxois (PBD \cite{PBD}) approach to DNA denaturation (\lq\lq melting")
is a representative example 
of such modeling. 
It was motivated by nonlinear lattice dynamics (\lq\lq phonon
softening”) and deals with the dynamics of each base pair in terms of a single 
scalar    
variable, which provides an effective measure 
of the spatial separation of the two bases.   
The combination of a local,  effective
intrapair interaction (modeled by a Morse potential and describing the combined effects of hydrogen bonding, stacking and solvent) 
and a nonlocal, nonlinear elastic interaction between nearest neighbor pairs (accounting for the effects of base-pair stacking)
leads to one of the simplest Hamiltonian models which exhibit an
exact thermodynamic phase transition in one dimension. Such a transition is predicted to occur in the case of
\lq\lq homogeneous" DNA, i.e. a polynucleotide of infinite extent \cite{TDP,JSP2001}. The heterogeneity
of natural DNA leads to a rounding of the transition, typically of the multistep
type \cite{CuleHwa}. Theoretical calculations with a revised set of model parameters have successfully matched observed melting profiles of long genomic chains with a 
rich internal structure, in solution \cite{NTh2010} as well as in fibers \cite{Wildes}. It would thus seem that the model has matured beyond \lq\lq proof of concept\rq\rq\- towards practical applicability.

Nonetheless, important discrepancies persist: First, the calculated melting entropy 
\cite{NTh2010} is less than half 
the calorimetrically measured transition entropy 
\cite{DelBlake,SantaLucia1996}. A similar discrepancy exists between model predictions \cite{NTh2019} and experimental data \cite{Essevaz-Roulet1997} regarding the magnitude of the force required for unzipping the two DNA strands.
Furthermore, 
rate constants of DNA hairpin dynamics calculated \cite{vanErp2015} with the same set of model parameters as in Ref. \cite{NTh2010}
are overestimated by several orders of magnitude compared to experiment; this has motivated further efforts \cite{vanErp2018} to improve on the model by adding a barrier to the on-site potential; 
however, this feature, although it improves agreement with experiment on the dynamics of short chains, does not seem to work for long chains. 
 
This work  has been motivated by the idea that the failure of the original PBD model to describe hairpin dynamics could be related to its incomplete description of equilibrium properties, namely the entropy and unzipping force discrepancies mentioned above, and that these discrepancies must be addressed first. In the next section, I will provide an alternative view of the microscopic, intra-basepair dynamics underlying the original PBD model.     
The proposed atomistic picture modifies the mesoscopic coarse-graining,
leading to the identification of {\em two} relevant degrees of freedom per base pair; to some extent, the two oscillators per base pair mimick the physical situation with its two or three hydrogen bonds.   
  Equivalently, this reduction of the double helix statistical mechanics can be thought of as a planar vector PBD chain with strongly anisotropic local and nonlocal nonlinearities, and 
 a concomitant  doubling of the calculated extensive thermodynamic quantities. 
Section III presents the calculation of the entropy in a standard case.   
Section IV presents results on the unzipping force. 
Section V addresses the questions raised by hairpin opening rates.
Concluding remarks are made in the final section and include an interpretation of  the equilibrium constant between open and closed states of a base pair during NMR imino proton exchange experiments \cite{Gueron1987}.

\section{Rethinking PBD}
\subsection{A toy model of the base pair plane}
\begin{figure}[h]
\vskip -.25truecm
\includegraphics[width=0.4\textwidth,height=0.4\textwidth]{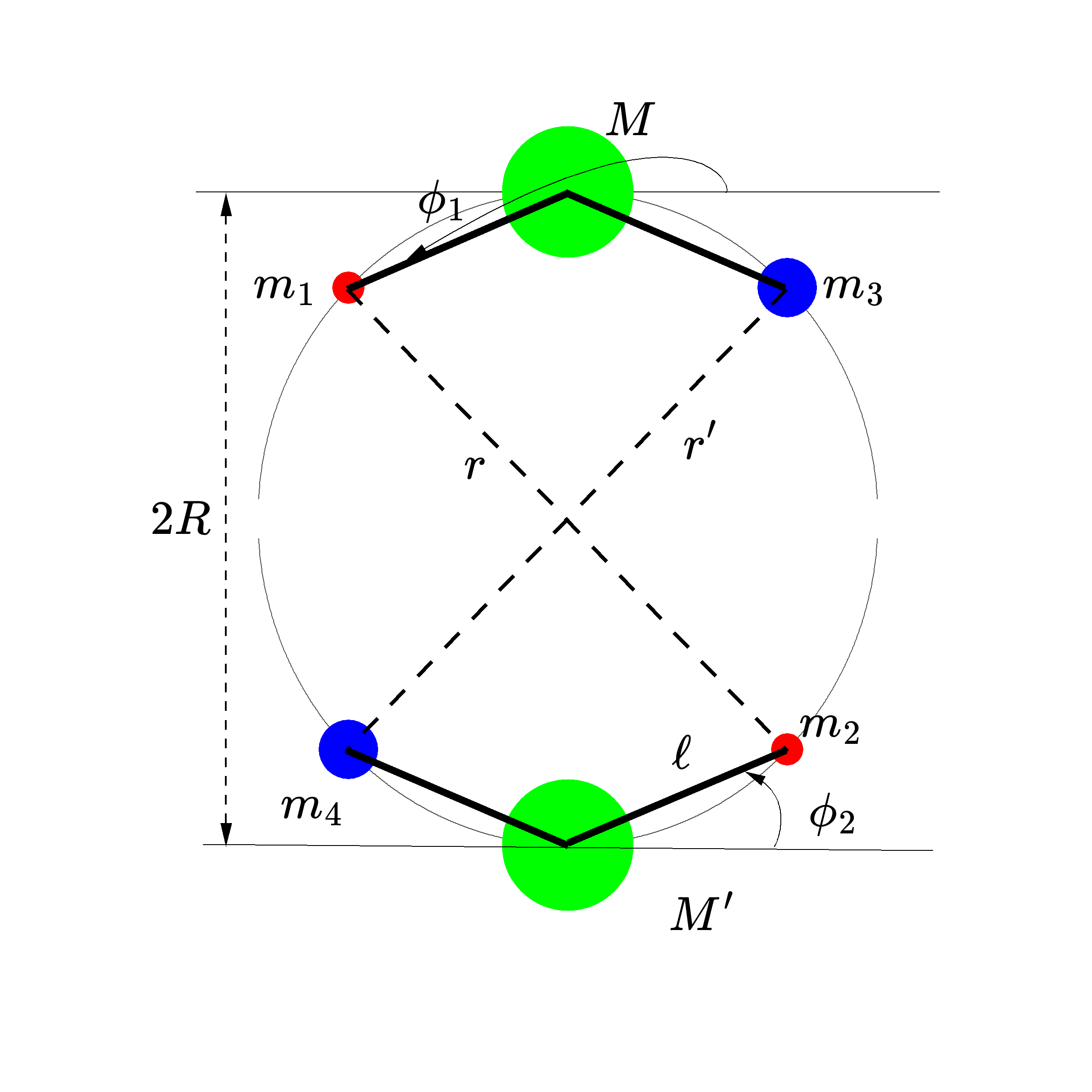}	
\vskip -0.5truecm
\caption{   
A toy model of a single base plane in the DNA double helix. The large full circles reperesent the core of each base, of mass $M$ and $M^\prime$, respectively, and are positioned at diametrically opposite points of a circle of radius $R$.
Attached to each base at a distance $l=2R \sin(\pi/8)$ are two point masses, one smaller, mimicking the carbonyl and exocyclic amino groups, and another approximating endocyclic nitrogen and its surrounding unit. The angle between the bond connecting the mass $m_i$ to its neighboring base core and the horizontal line is $\phi_i$; at equilibrium, $ \phi_2^{(0)} = \phi_1^{(0)} - \pi  = \pi/8 $.
The dashed lines represent the effective couplings between the complementary bases; they are perpendicular to each other at equilibrium.
Since the  mass ratios $m_i/M$, $m_i/M^\prime \ll 1$  the core parts of the bases are approximately fixed in space. 
At equilibrium, all 4 mass points are on the perimeter of the circle and the distances $r$ and $r^\prime$ are equal to the circle's diameter $2R$.
} 
\label{fig:BPP}
\end{figure}

The model is illustrated in Fig. \ref{fig:BPP}; each of the two complementary bases consists of a main part and two attachments of smaller mass held rigidly at a distance  $l$ from the center of the main part. 
The direction of the bond between main part and attachment is free to vary, but is assumed to remain on the base plane.
The attachments with the lowest mass $m_1, m_2$ are meant to mimick the carbonyl and the exocyclic amino groups, which are responsible for the hydrogen bonding between complementary bases and therefore typically have a mass of about $16$ amu, far smaller than the remaining  mass $M$ of a base, which is of the order of 300 amu. The attachments with the intermediate masses $m_3, m_4$ are meant as approximations to endocyclic nitrogen atoms and their surrounding units. Again, it is assumed that these masses too are substantially lower than $M$.
This means that the dynamics of  the entire planar system can be well approximated by the dynamics of the 4 point masses $m_i, (i=1,2,3,4)$, while the main parts of the bases can be thought of as being fixed in space. 
The dashed lines represent the effective coupling between complementary bases; they are perpendicular to each other at equilibrium;



I will use a Cartesian coordinate system with the center of the circle (Fig. \ref{fig:BPP}) as the origin. The positions of the 4 mass points $m_i$ can be described in terms of the direction angles $\phi_i$ relative to the horizontal axis. Noting that the motion of the masses 1 and 2 is decoupled from that of 3 and 4, I can write the Lagrangian of the former as
\begin{equation}
{\cal L} = \frac{\ell^2}{2} \sum_{i=1}^2  m_i {\dot\phi_{i}}^{2}   - V(r-r_0)   \quad,
\label{eq:Lagr}
\end{equation}
where 
\begin{equation}
(\frac{r}{2R})^2  =  1 + (\frac{\ell}{R})^2  \sin^2(\frac{\phi_1-\phi_2}{2}) +       \frac{\ell}{R}(\sin \phi_1 - \sin \phi_2)                          \quad,
\label{eq:rphi}
\end{equation}
$V$ is the interaction potential between the two point masses, 
$r$ their separation distance, and $r_0=2R$ the latter's equilibrium value.
As in the PBD model, I will use a Morse potential with depth and width parameters $D$ and $1/\alpha$, respectively,
\[
V(\eta) = D (1-e^{-\alpha \eta}  )^2
\]
and its harmonic limit
\[
V_h (\eta) = \frac{1}{2} K \eta^2  \quad,
\]
where $K=2D\alpha^2$. 

The equations of motion corresponding to (\ref{eq:Lagr}) are
\begin{equation}
m_i \ell^2 \ddot{\phi}_i = - \frac {\partial V}{\partial \phi_i}  \quad , i=1,2 \quad.
\label{eq:eqmot}
\end{equation}
At equilibrium, $r=r_0=2R$,
\[
\phi_2 = \phi_2^{(0)} = \phi_1^{(0)} - \pi = \arcsin \frac{l}{2R}  = \pi/8 \quad.
\]
A completely analogous  set of equations describes the dynamics of the subsystem composed of the masses 3 and 4. 

In the limit of small oscillations around equilibrium,  $\delta \phi_i = \phi_i-\phi^{(0)}_i = \delta \phi_i^{(0)} e^{i \omega t} $, the 
equations of motion (\ref{eq:eqmot}) lead to two distinct eigenmodes. One eigenmode has a zero eigenvalue and its  eigenvector components satisfy the relationship
\[
 \delta \phi_1^{(0)}+ \delta \phi_2^{(0)}=0    \quad.
\]

The other eigenmode has a nonzero oscillation frequency
\begin{equation}
\omega^2  = K \cos^2(\pi/8) (\frac{1}{m_1}+\frac{1}{m_2})   
\label{eq:frequency}
\end{equation}
and the corresponding unnormalized eigenvector satisfies
\[
\frac{m_1}{m_2} \delta \phi_1^{(0)} -  \delta \phi_2^{(0)}=0    \quad.
\]
It is straightforward to see that the zero frequency mode reflects the local invariance of the Hamiltonian to small, equal and opposite rotations of the bonds $(Mm_1, M^{\prime}m_2)$; the latter can be seen to leave the distance $r(\phi_1,\phi_2)$ unchanged. Any nontrivial dynamics of the subsystem $\{m_1,m_2\}$ originates in the 
nonzero frquency eigenmode. This in turn, except for the geometrical factor, is just the dynamics of a PBD \lq\lq diatomic pair\rq\rq coupled by the interaction potential $V$.

Identical arguments hold for the subsystem $\{M^\prime ,m_3,m_4\}$, the only difference being that the oscillator frequency (\ref{eq:frequency}) will now be lower on account of the higher masses.
\subsection{A two-chain version of the PBD model}
The toy model presented above may serve as an \lq\lq atomistic foundation\rq\rq of a variant of  the original PBD model. 
%
%
In this variant,
there is one more independent oscillator per base pair, originating in the motion of
the subsystem $\{M^\prime ,m_3,m_4\}$. Note that while this bears some correspondence to the physical situation of poly(AT) chains with the two hydrogen bonds per base pair, the model will have to deal (as in the original PBD case) with both AT and GC pairs parametrically, i.e. by adjusting the parameters of the Morse potential to stronger GC  bonding while taking account of the combined effects of hydrogen bonds, stacking and solvent.

The total configurational energy of the proposed system will be composed of two functionally identical, independent, -and therefore separable- parts
\[
H^{P}_{tot} = H^{P}(\{\eta_i\}) + H^{P}(\{\eta_{i}^{\prime}\})  \quad,
\]
where 
$\eta_i$, $\eta_{i}^{\prime}$, 
denote displacements  of two independent one-dimensional oscillators along mutually orthogonal directions in the base plane,
\[
H^{P}(\{\eta_i\}) = \sum_{i=1}^{N-1} W(\eta_{i},\eta_{i+1}) +  \sum_{i=1}^{N} V(\eta_{i} )   
\]
is the potential energy part of the PBD Hamiltonian, and 
\[
W(\eta_1,\eta_2)   =\frac{1}{2} k [1 + \rho e^{-b(\eta_1+\eta_2)}  ] (\eta_1-\eta_2)^2
\]
represents  the nonlinear stacking interaction, parametrized by the stacking range $1/b$, the residual stacking strength $k$ of the high-temperature state, and the ratio $\rho\gg 1$ of the stacking strengths of double vs. single-stranded DNA. It
should be noted that, as in the original PBD formulation, the on-site potential $V(\eta)$  lumps together effects of hydrogen bonding, stacking and solvent. Furthermore, the fact that the particles of the second chain have a higher mass changes the kinetic energy but does not enter the configurational energy. The chains are, insofar as configurational partition functions are concerned, identical.

The configurational partition function $Z_N^{tot}$ of the total system composed of $N$ base pairs will thus be equal to the square of the original partition function $Z_N$ as used in e.g. \cite{PBD} or \cite{NTh2010}. Explicitly, for $N\gg 1$ and homogeneous systems
\[
Z_N^{tot} = Z_N^2   \quad,
\]
where
\[
Z_N \sim \Lambda_0^{N-1}I_0^2
\]
\begin{equation}
I_0 = \int_{-\infty}^{\infty} d\eta\> e^{- V(\eta)/(2k_B T) }\psi_0(\eta)
\label{eq:I0}
\end{equation}
and $\Lambda_0$, $\psi_0$ are respectively the largest eigenvalue and the corresponding eigenfunction of the transfer integral (TI)  operator, $T$ is the absolute temperature and $k_B$ the Boltzmann constant. 
Details of notation of TI thermodynamics and generalization to inhomogeneous systems are given in \cite{NTh2010,NTh2011,NTh2019}.

Hence, all extensive thermodynamic quantities, e.g. enthalpies and entropies and, in particular, the melting entropy as calculated in \cite{NTh2010}, should be multiplied by a factor of 2. Intensive quantities, e.g. the melting fraction, remain the same.

\subsection{Equivalence with a planar vector model}
The geometry of the base plane (cf Fig.\ref{fig:BPP}) suggests that we may treat the pair $(\eta_i, \eta^{\prime}_i)$ as a two-dimensional vector. The model may then be thought of as a single discrete planar vector chain with anisotropic local nonlinearity (the decoupled Morse potentials in two independent directions on the base plane) and an anisotropic nonlocal nonlinear interaction (the nonlinear part of the base stacking term acting in both independent directions). 
In this version of the theory, because of the particular form of the anisotropy, the dynamics in the two directions becomes separable and the partition function factorizes.

A planar vector model with strong anisotropy, both local and nonlocal, appears {\em a posteriori} better adapted to the symmetry of  DNA, independently of the derivation presented here in terms of a particular \lq\lq toy model\rq\rq .  This alternative formulation might  also be useful as a starting point for other, more realistic, less anisotropic vector models of DNA melting. Regarding the local anisotropy, the approach is in fact not new; one-dimensional Morse potentials have been succesfully used e.g. to describe melting of hydrogen bonds in the context of self-consistent phonon theory \cite{Proho84}. Deviations from the extreme anisotropic case of the present model are more likely to originate in the nonlocal coupling; however,  judging from the results presented in the next sections, I expect cross-correlation effects arising from them to be weak.

In this work, I will refer, equivalently, to either a \lq\lq two-chain\rq\rq\- or a \lq\lq vector\rq\rq\- generalization of the PBD model.

\section{The PBD entropy of melting}
\subsection{Numerical estimate in the general, heterogeneous case}
The method for calculating the thermodynamic functions is described in \cite{NTh2010}. The model parameters, as used in that work, are
$b=0.2 A^{-1}$, $k=0.45 \times 10^{-3} eV/A^2$, $\rho=50$, $\alpha_{GC}=6.9 A^{-1}, \alpha_{AT}=4.2 A^{-1}$, and a cutoff $\eta_c=2 A$, beyond which a PBD base pair is considered open; furthermore, the values
$D_{GC}=0.1655 \> eV, \> D_{AT}=0.1255 \> eV$, which were used to fit the melting profile of PBR322 (Eco RI cut) \cite{DelBlake}, serve as reference values at salt concentration 0.075M.



\begin{figure}
\includegraphics[width=0.4\textwidth]{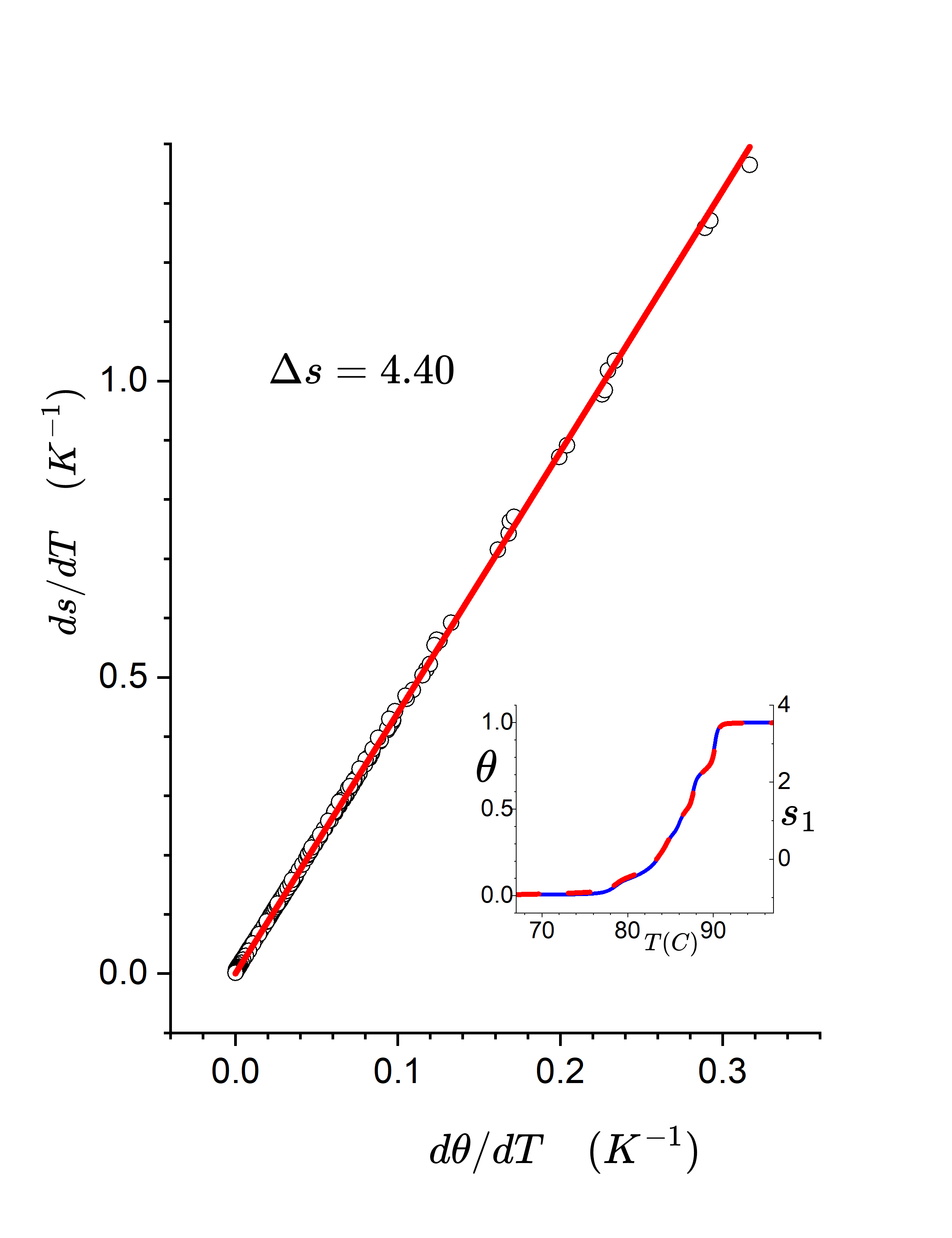}
\vskip -0.2truecm
\caption{Melting entropy for a single PBD chain. The onset shows that the curves for the entropy $s_1$ (right y-axis, red, dashed) and the melting fraction (left y-axis, solid, blue) $\theta$ coincide when appropriately rescaled. 
The main plot shows the temperature derivative of the anharmonic part of the entropy vs the temperature derivative of the melting fraction (open circles). 
The linear fit to the data through the origin yields the slope $\Delta s^*$. 
The absence of scattering in the data points (which originate in different regions of the melting profile) confirms the validity of the Ansatz (\ref{eq:ds}) and, hence, that the melting entropy of a base pair is largely independent of chemical composition and/or position in the sequence \cite{DelBlake}.
}
\label{fig:MeltingEntropy}
\end{figure}
The melting fraction and the entropy per site, calculated using the above set of parameters and the full genomic sequence of PBR322 \cite{NTh2010}, are shown in the onset of Fig. \ref{fig:MeltingEntropy}; appropriately rescaled, the two curves are essentially identical.
In order to obtain a numerical estimate of the melting entropy I make use of 
the {\em Ansatz} 
\begin{equation}
\frac{ds_{1}(T)}{dT} = \Delta s^{*}\frac{d\theta}{dT}
\label{eq:ds}
\end{equation}
where $s_{1}(T)$ is the entropy per site at temperature $T$ in units of $k_B$, and $d\theta/dT$ the temperature derivative of the melting fraction, both computed for a scalar chain, it is possible to obtain a theoretical estimate  of  the melting entropy per site $\Delta s^{*}=4.40\>$. The procedure is shown in Fig. \ref{fig:MeltingEntropy}.
Since the numerics is done for a scalar PBD chain, the end result for the vector model is 
\begin{equation}
 2 \times \Delta s^{*} = 8.8   \quad .
\end{equation}

\subsection{Elastic entropy} 
Neither the PBD model in its original Hamiltonian form nor its vector generalization can describe entropic elasticity.  Each base pair may dissociate, in the sense that the base pair separation coordinates exceed a certain value, yet, strictly, continues to reside on the base plane and is not free to sample 3-dimensional space. The PBD \lq\lq molten\rq\rq\- state can be thought of as a  longitudinally \lq\lq straight\rq\rq\- ladder  with very wide steps and no overall bending. The change in entropy which originates in the different embedding of such a stiff polymer with very small orientational entropy vs  two soft single-stranded polymer chains in 3d-space can in fact be estimated separately, as follows.


A straightforward way to compute the elastic entropy of a polymer composed of $N$ monomers, monomer distance $a$ and persistence length $\lambda$ is to approximate the polymer by a freely-jointed chain (FJC) consisting of $N/(2\lambda /a)$
stiff segments, each  of which has a length $2 \lambda$ (Kuhn length). Each segment contributes a factor $4\pi$ to the conformational partition function
\[
\Omega = (4\pi)^{aN/(2\lambda)}  \quad,
\]
leading to an elastic entropy of the polymer, in units of $k_B$,
\[
S_{el}=  \ln \Omega = \ln (4\pi)^{aN/(2\lambda)} \quad.
\]
The orientational (elastic) melting entropy of DNA per base pair can then be estimated as the difference
\begin{equation}
  \Delta s_{el} \equiv  \frac{\Delta S_{el}}{N} = \ln(4\pi) \{ 2\frac{a'}{2\lambda^\prime} -\frac{a}{2\lambda} \}
\label{eq:meltentr}
\end{equation}
where $a,a^\prime,\lambda, \lambda^\prime$ are, respectively, the monomer and persistence lengths of double- and single stranded chains,
and the factor of 2 reflects the two separate strands of the denatured state.
Using standard values $a=0.34$ nm, $a^\prime = 0.63$ nm \cite{Murphy2004}, $\lambda = 53.4$ nm \cite{Bustamante1994}, $\lambda^\prime=1.62$ nm \cite{Ritort2014}, results in a value $ \Delta s_{el} =0.98$ per base pair. 

It should be noted here that the second term in (\ref{eq:meltentr}) is very small, as expected; the elastic entropy of double-stranded DNA is essentially zero. The main contribution to $\Delta s_{el}$ comes from the first term and is sensitive to the choice of parameters. I have chosen the carefully averaged SAXS value of the monomer distance reported in \cite{Murphy2004}. The value of  $\lambda^{\prime}$ used was  interpolated 
from results of \cite{Ritort2014} for a salt concentration of $0.075$ M, consistent with the data used in the previous section; lower values were reported at higher salt concentrations and vice versa. 

Combining the above estimate of 3-d elastic entropy with the result of the previous section, I obtain a value of
\begin{equation}
  \Delta s = 2 \times \Delta s^{*} + \Delta s_{el} = 9.78 \quad.
\label{eq:meltentrtot}
\end{equation}
This is roughly in line with experiments
 reporting a melting entropy averaged over many calorimetric experiments on long polymers, $\Delta s_{exp} =12.5 \pm 0.9 $  \cite{DelBlake}, and the somewhat lower estimate, $\Delta s_{exp} =11.0$ of \cite{SantaLucia1996}, based on a systematic analysis of oligomer data.

\section{Unzipping}
\begin{figure}
\includegraphics[width=0.4\textwidth]{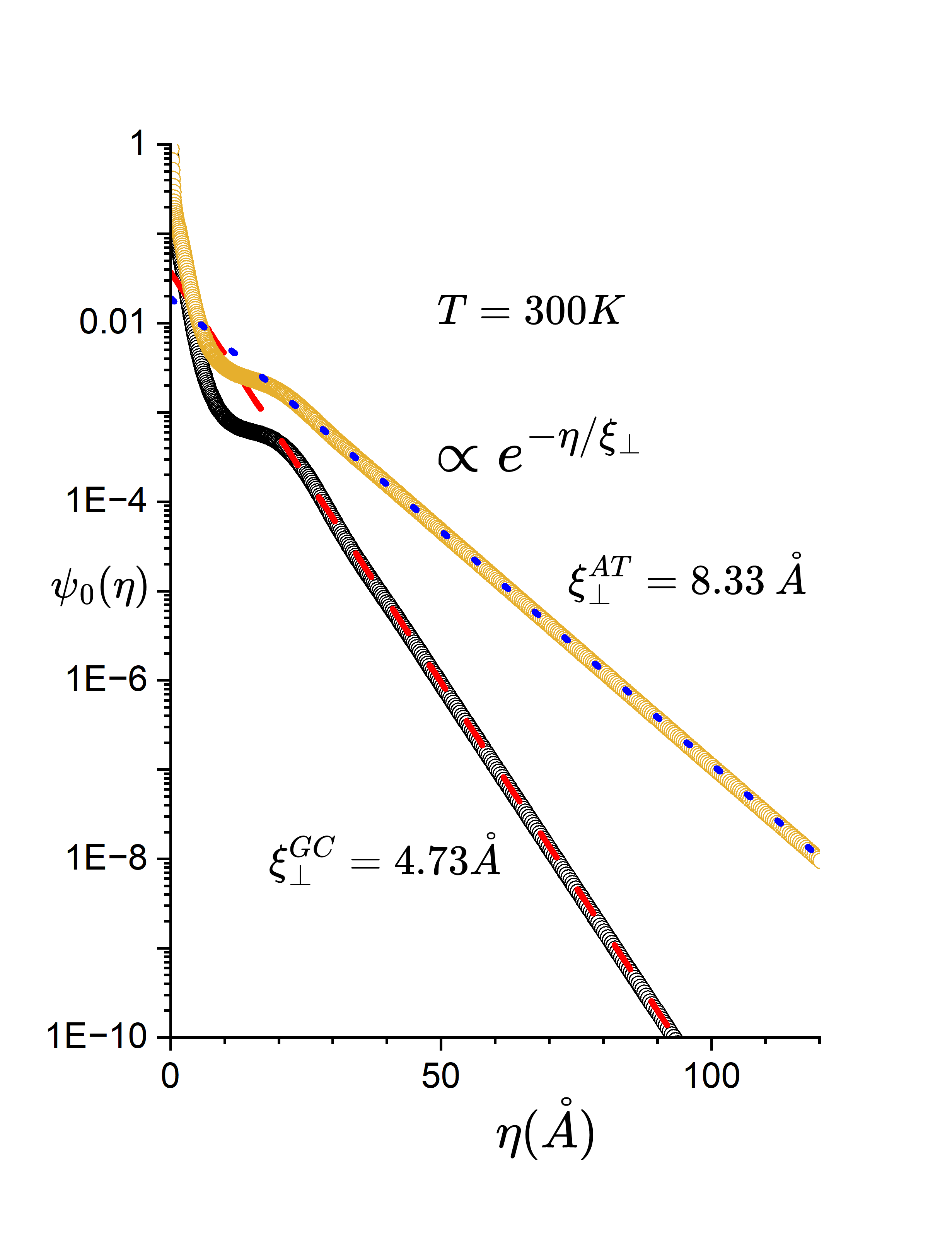}
\vskip -0.2truecm
\caption{The eigenfunction $\psi_0$ in the homogeneous case. Note the purely exponential asymptotics which allow the extraction of $\xi_\perp$ for GC and AT. Parameters are the same as in Section III.
}
\label{fig:GSeigenf}
\end{figure}

Calculating the transverse force required to separate the two DNA strands is quite straightforward in the homogeneous limit within the PBD model \cite{NTh2019}. In the presence of such a force, the Hamiltonian is modified by a term $-f \eta_N$, where $\eta_N$ is the displacement coordinate of the last base pair. This modifies one of the two end integrals $I_0$ in Eq. \ref{eq:I0}, introducing an extra factor $e^{ f \eta/(k_B T)}$ in the integrand. Now it is known from the continuum limit of the PBD model \cite{JSP2001} that the eigenfunction $\psi_0$ decays exponentially at large values of its argument. This property survives in the discrete case as well, as shown in Fig. \ref{fig:GSeigenf}, i.e. $\psi_0(\eta)\propto e^{-\eta/\xi_\perp}$, where $\xi_\perp$ is a characteristic length scale. As the force increases, an instability (divergence of $I_0$) occurs at a value
\[
f^*= \frac{k_B T}{\xi_\perp}   \quad.
\]
Within the vector model introduced here, the unzipping force is equal to $2f^*$, since half of the physical external force must be applied to each chain. The result, for the parameters of Section III (corresponding to a salt concentration of 75 mM) is an unzipping force of 10 pN for AT and 17.6 pN for GC at 300 K, i.e. an average of 13.8 pN. In comparison, Ref. \cite{Essevaz-Roulet1997} reports 13 pN at 60\% GC content and 150mM salt concentration, whereas separately measured unzipping forces of  $9\pm3$ pN and $20\pm3$ pN have been reported for AT and GC, respectively\cite{Gaub1999}, also at 150 mM salt concentration.

\section{Hairpin opening rates}
Single strands of DNA with complementary ends may close on themselves to form hairpin-like structures. The complementary ends form the stem, which is effectively a double-helical oligomer, while the rest of the chain arranges itself in a loop-like fashion. Raising the temperature results in dissociation of the hairpin structure via melting of the stem. The melting curves in the limiting case of a short stem and an intermediate-size T-loop \cite{Libchaber2000} can be well understood \cite{NTh2019} in terms of the stem nearest-neighbor enthalpies and entropies given in \cite{SantaLucia1997} and a self-avoiding walk polymer model of the loop.  The complete thermodynamic and kinetic problem is a complex one, involving both DNA and polymer statistical mechanics, diffusion properties of ss-DNA, as well as indermediate states formed during trapping by mismatches in the process of folding \cite{Ansari2001,Errami2007}. In this Section I will consider measurable quantities for which such complexities are either absent or they cancel out, e.g. in ratios of reaction rates.  In such simple cases hairpin opening occurs by melting \cite{Libchaber1998, Libchaber2000} or \lq\lq unzipping\rq\rq \cite{Zocchi2007, Woodside2006} of the stem; the barrier for the opening transition is 
determined by the stem's free energy difference relative to the open state \cite{Zocchi2007}, which contains both entropic and enthalpic contributions, i.e. the opening rate is 
\begin{equation}
k_{op} = \gamma_0  \> e^{\Delta S_{stem}/k_B}    e^{-\Delta H_{stem}/(k_B T)}  \quad,
\label{eq:opening}
\end{equation}
where $\gamma_0$ is controlled by the diffusion rate of ss-DNA and can be determined by the measured closing rate \cite{Libchaber1998} to be of the order of $15 \> \mu sec^{-1}$.

In the case of the 31-base sequence $5^{'}-CCCAA-(T)_{21}-TTGGG-3^{'}$, the transition enthalpy has been determined to be equal
to $32$ Kcal/mol  \cite{Libchaber1998}. This value can be understood in terms of the nearest-neighbor DNA melting enthalpies \cite{SantaLucia1997},
\begin{eqnarray}
\nonumber
\Delta H_{stem}=\Delta H_{C \circ G}^{init}+ 2 \Delta H_{CC/GG} + \Delta H_{CA/GT} \\
\nonumber
+\Delta H_{AA/TT}+\Delta H_{A \circ T}^{init}\\
\nonumber
=-0.1 + 2\times 8 + 8.5 + 7.9  -2.3 = 30 \> Kcal/mol,
\end{eqnarray}
where initiation values have been used for the stem end points.

\begin{figure}
\resizebox{0.5\textwidth}{!}
{\includegraphics[width=0.26\textwidth]{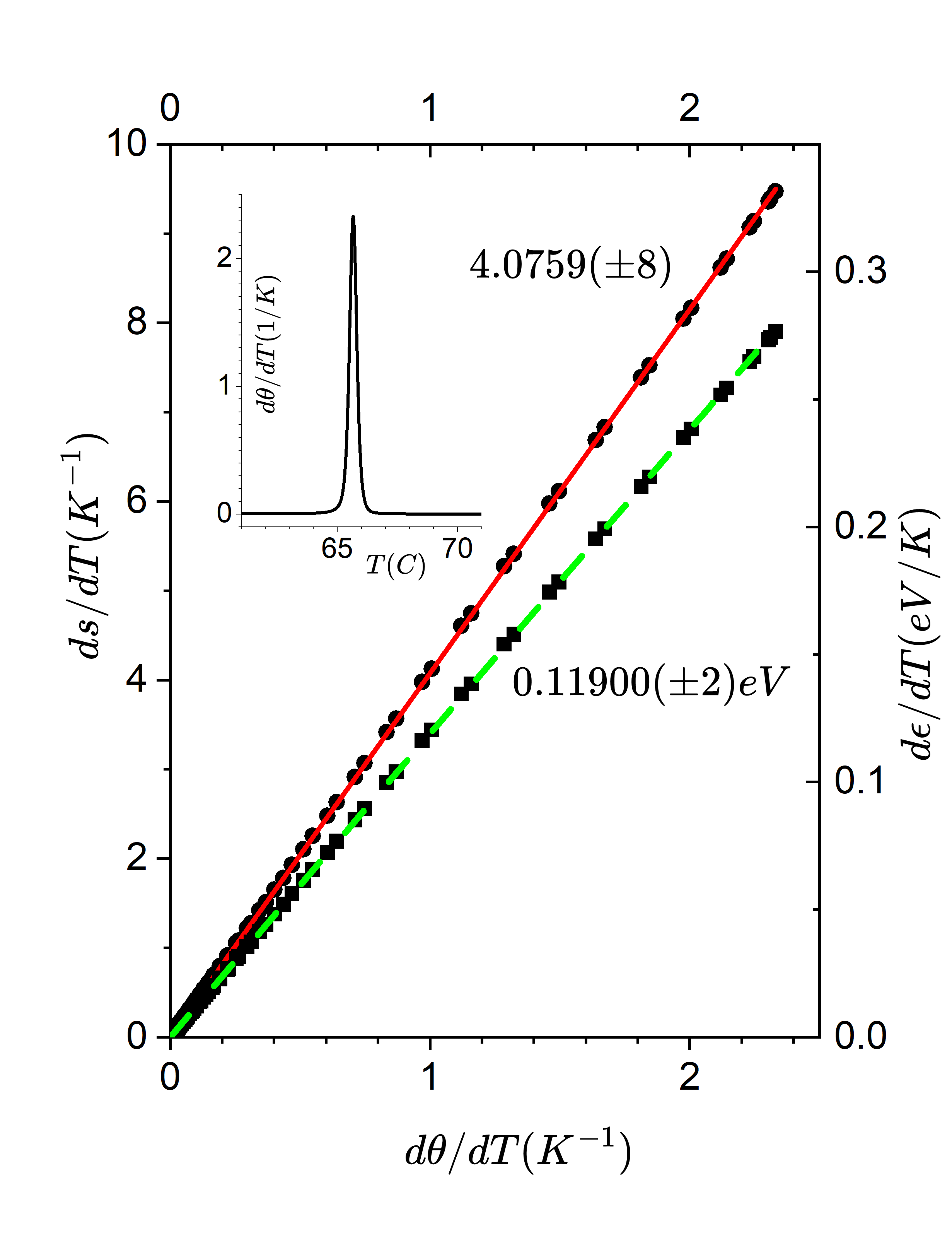}	
\includegraphics[width=0.26\textwidth]{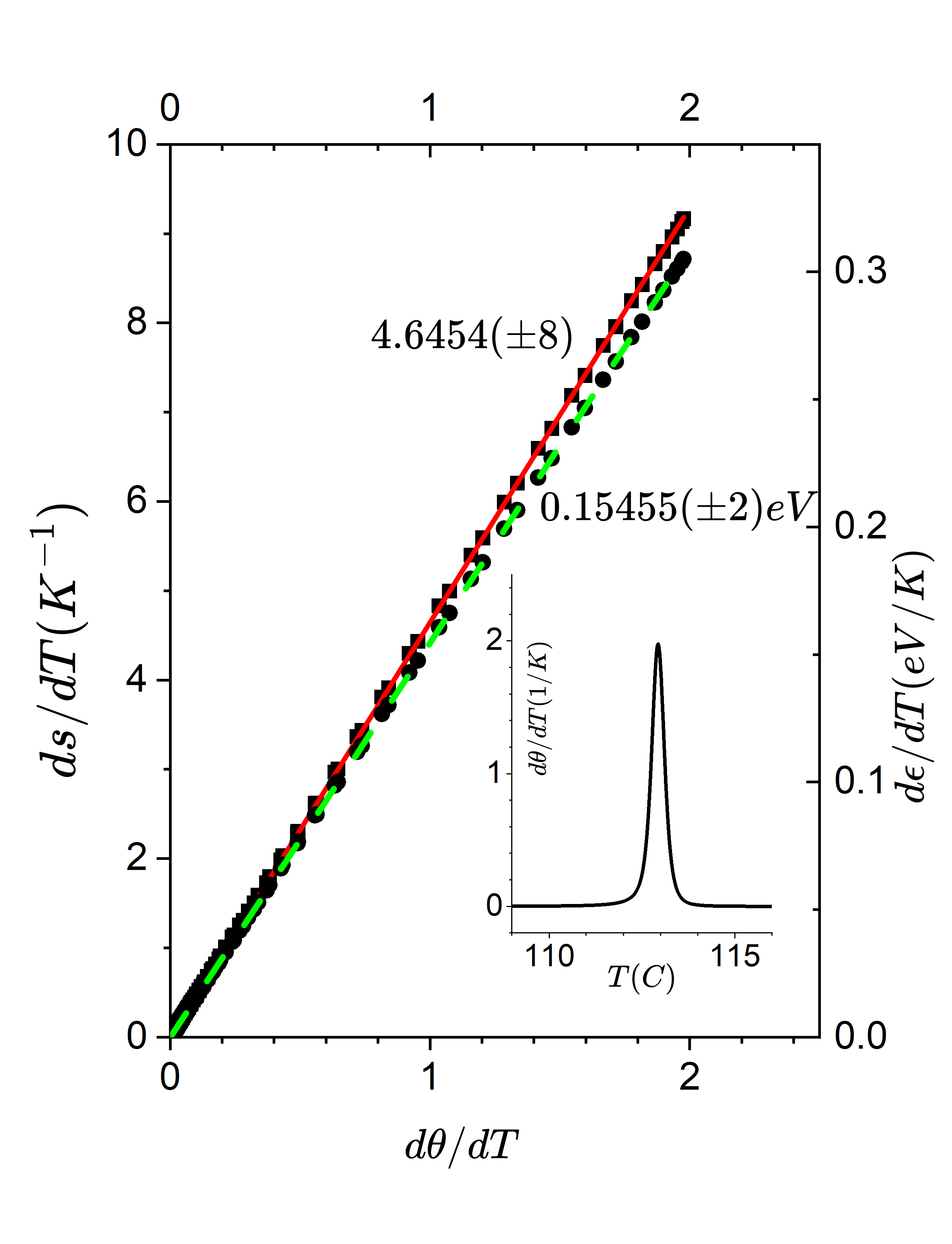}}
\vskip -0.2truecm
\caption{Extraction of single chain PBD (N=1000) average energies and entropies per site (cf Fig. 1);
{\em left panel}: pure AT: $ds/dT$ as a function of the temperature derivative of the melting fraction (squares, left y-axis);
$d\epsilon/dT$ as a function of the temperature derivative of the melting fraction (circles, right y-axis); the lines represent least square fits and their slopes provide estimates for melting entropy and energy, respectively; the onset shows the temperature derivative of the melting fraction as a function of 
temperature. {\em right panel}: same for pure GC. The Morse depth parameters used, $D_{AT}=0.1291\>eV$, $D_{GC}=0.1681$, correspond to a salt concentration 0.2M.
}
\label{fig:GCandATthermo}
\end{figure}

Alternatively, in the framework of the PBD model, the transition enthalpy can be estimated by assigning enthalpies to single base pairs. The values which enter such estimates can be obtained by computing the thermodynamics of long (N=1000) sequences of pure AT or pure GC. I use the parameters introduced in Section III, modified for salt concentration 0.2 M according to \cite{NTh2010}. This results in a slight shift of the Morse potential depths, $D_{GC}= 0.1291 \>eV   $ and $D_{GC}= 0.1681\>eV $, respectively.
The relevant results of TI thermodynamics are displayed in Fig. \ref{fig:GCandATthermo}, from which one extracts transition enthalpies 
$\epsilon^{(0)}_{GC} = 0.1545 \> eV$, $\epsilon^{(0)}_{AT}=0.1190 \> eV$ and transition entropies $s^{(0)}_{GC} = 4.6454$, $s^{(0)}_{AT}=4.0759$, respectively.  Including the elastic entropy contribution, this results in total entropies per base pair  $s_{GC} = 10.27$, $s_{AT}=9.13$. The corresponding melting temperatures are $T^{m}_{GC}=\epsilon^{(0)}_{GC}/s^{(0)}_{GC}=386.10 \>K$ and $T^{m}_{AT}=\epsilon^{(0)}_{AT}/s^{(0)}_{AT}=338.82 \>K$ and agree with those determined directly from the profiles of Fig. \ref{fig:GCandATthermo} within 0.03 K. 

In order to maintain consistency of the elastic-entropy-corrected thermodynamics with the melting profiles it is necessary to correct the single chain GC and TA enthalpies by amounts $T^{m}_{GC}\Delta s_{el}/2$ and $T^{m}_{AT}\Delta s_{el}/2$ respectively. The resulting  total transition enthalpies per base pair for the vector model are  $\epsilon_{GC} = 0.3417 \> eV$, $\epsilon_{AT}=0.2666  \> eV$, respectively. I note in passing that these values are within 3\% of the respective Morse potential depths multiplied by a factor of 2. 

The total enthalpy of the stem can now be estimated as
\begin{eqnarray}
\nonumber
\Delta H_{stem} =  3\epsilon_{GC} + 2\epsilon_{AT}\\
\nonumber
 = 1.5583\> eV =35.9 \> Kcal/mol \quad,
\end{eqnarray}
or 60.7 in units of $k_B T$ at 25 C,
about 12\% higher than the experimentally determined value. This is consistent with the difference between the enthalpies derived in this work and the somewhat lower effective enthalpy parameters 0.2910 and 0.2465, for GC and AT respectively \cite{vanErp2015}, which fit the experimental data of \cite{Libchaber1998}. 


As a second hairpin example, where now entropic stem effects are important, I will consider the case of two otherwise identical hairpins, where the 10-basepair long stems differ only at a single site, i.e. GAAGAGGGAG vs GAAGAGGGGG. In this case the ratio of the opening rates has been measured at $25$ C to be $5.6 \pm 1.5$ \cite{Zocchi2007}. The substitution of a single GC by an AT pair produces, in general, both an energy and an entropy difference. Using the nearest neighbor model with the standard parameter set \cite{SantaLucia1997} results in zero enthalpic change and a $\Delta S^{*}$=3.4 cal/K/mol. The opening rate ratio is estimated from 
(\ref{eq:opening})
as $e^{\Delta S^{*}/k_B}=5.53$, in excellent agreement with experiment.

In the vector PBD case, the rate ratio will be
\begin{equation}
e^{(s_{AT}-s_{GC})}e^{-(\epsilon_{AT}-\epsilon_{GC})/(k_B T)} = e^{-1.14+2.92}=5.93 \quad,
\end{equation}
also in line with experiment.

In order to address the discrepancy of PBD opening rates by several orders of magnitude compared to experiment, a problem raised in Ref. \cite{vanErp2015}, one must return to (\ref{eq:opening}) and consider the entropic term as well. In the case of the 5-stem hairpin considered above, the stem's  entropic contribution within the vector PBD framework is
\[ 
\Delta S_{stem}/k_B=  3s_{GC} + 2s_{AT} =  49.1 \quad.
\]
At a temperature of 25 C, this results in a free energy difference $60.7-49.1=11.6$ in units of $k_B T$, and an opening rate $k_{op}/\gamma_0= e^{-11.6}=9.17\times 10^{-6}$.
Had I used the single-chain PBD model, the result would have been $5.2 \times 10^{-3}$, i.e. two and a half orders of magnitude higher; this is exactly the order of  discrepancy between PBD and experimental values exhibited in Fig. 6 of Ref. \cite{vanErp2015}.  

As the stem length increases to $N_{stem}=30$ base pairs, the discrepancy between single-chain PBD opening rates predictions and experimental values rises up to ca. 17 orders of magnitude (cf. Fig. 4 of Ref. \cite{vanErp2015}). 
At 50\% GC content and 25 C temperature, the average values of vector PBD melting energy and entropy per site are $\epsilon = 0.3042\> eV=11.95 \>k_B T$ and $s=9.70$ respectively. The opening rate is then expected to be proportional to $b^{N_{stem}}$ where $b= \exp\{-\epsilon/(k_B T)+s\}=\exp(-11.95+9.70)=\exp(-2.25)=0.105$. The corresponding number extracted from experiment \cite{Woodside2006} in the analysis of \cite{vanErp2015} is $b=0.095$, i.e. lower by approximately 10\%. 
For a stem length equal to 30, this will result in a discrepancy of $(0.105/0.095)^{30}=20$, i.e. the opening rate calculated within the  vector PBD model will still be higher than the observed one by a factor of 20. While not negligible, this is certainly a long way from the 17 orders of magnitude discrepancy generated by the single-chain PBD model.

\section{Discussion}
This paper has argued that a reductionist model of DNA melting according to the logic of Ref. \cite{PBD} can follow naturally from a microscopic picture of the base-pair plane of the double helix in which carbonyl and exocyclic as well as endocyclic amino groups appear explicitly. The novelty of the particular reductionist scheme is that it leads to two independent, nonlinear oscillators per base pair, which, in the presence of the stacking interaction,  generate two independent chains of coupled nonlinear oscillators, thus doubling the magnitudes of extensive thermodynamic quantities while leaving melting profiles as computed in e.g. \cite{NTh2010} unchanged. The analysis of diverse types of data, e.g. melting entropy, unzipping forces and hairpin opening rates presents rather compelling evidence in favor of the vector PBD model introduced in  this work. However, it should be noted that this type of analysis simply shows a very complete {\em thermodynamic equivalence} between a Hamiltonian model with a particular set of 7 parameters and the nearest-neighbor model with its set of 10 enthalpies and 10 entropies. There is as yet no specific experimental evidence in favor of the microscopic physics underlying the PBD model, either in the vector or in its original scalar version. On a somewhat speculative vein, one might therefore consider the following problem:
 

NMR experiments with DNA dodecamers using imino proton exchange suggest that DNA base pairs open spontaneously at room temperatures, one at a time \cite{Gueron1987}. The equilibrium constant between open and closed states of a base pair has been measured to lie between $10^{-5}-10^{-6}$, the lower values corresponding to GC and the higher to AT pairs. For comparison, a PBD-based long chain calculation performed with the parameters of Section III at room temperatures yields melting fraction values  $\theta = 0.7\times 10^{-3}$ (GC) and $2.5\times 10^{-3}$ (AT), which are two to three orders of magnitude larger. Setting for a moment aside the question of whether long chain results can be applied to small systems (the example of hairpins shows they - approximately - can), one may pause to examine the measurement process. The typical experimental situation in calorimetry or UV absorption, is that the disruption of a single hydrogen bond (as opposed to a full opening of a base pair) at any number of sites is detected as proportionately contributing to the melting fraction. Similarly, the melting fraction computed in the PBD model is the fraction of sites where the spatial separation coordinate exceeds a certain value; adding an uncorrelated identical chain does not change the computed value of $\theta$. In contrast, imino proton exchange requires a fully open state of the base pair, i.e. in the context of the present paper, a simultaneously open $i$th site on {\em both} chains (or, in the vector language, both independent components of the vector order parameter exceeding a certain value). The probability for such an event to occur is $\theta^2$, which at room temperatures has a much lower value, namely, $0.5\times 10^{-6}$ and $6\times 10^{-6}$ for pure GC and AT respectively. These values are close to the equilibrium constants determined in \cite{Gueron1987}. 

In this context, it is interesting to note that, in order to obtain such low values of the open state probability with the same parameters in a standard PBD calculation, a much higher value of the cutoff $\eta_c = 5 A$ would be needed \cite{NTh2010}; that result already indicated that the open state of imino proton exchange was probably not identical with what calorimetry or UV absorption detects. The model presented here clarifies this difference in a more satisfactory manner. 

The removal of the serious discrepancy between theory and experiment on the opening rate of hairpins has an important, perhaps not immediately obvious, consequence regarding further research. Usage of the PBD model in mesoscopic molecular dynamics simulations of dynamical properties, as has been done in \cite{vanErp2015} has been explicitly validated; as a result, the interpretation of PBD simulation data on dynamical properties has become less speculative. It would be interesting to explore this e.g., in order to provide an up-to-date theoretical estimate of the average lifetime of a base pair \cite{Gueron1987}.
\vspace{0.5cm}
 
\section{acknowledgement}

I acknowledge thoughtful comments and critical questions raised by both referees. They have resulted in a substantially improved version of this paper.


\end{document}